# Title: Dual-microcavity narrow-linewidth Brillouin laser

**Authors:** W. Loh[1], A. A. S. Green[1], F. N. Baynes[1], D. C. Cole[1], F. J. Quinlan[1], H. Lee[2], K. J. Vahala[2], S. B. Papp[1], and S. A. Diddams[1]

**Affiliations:**

[1]Time and Frequency Division, National Institute of Standards and Technology, Boulder, CO 80305, USA.

[2]T. J. Watson Laboratory of Applied Physics, California Institute of Technology, Pasadena, CA 91125, USA.

**Abstract**:

Ultralow noise, yet tunable lasers are a revolutionary tool in precision spectroscopy, displacement measurements at the standard quantum limit, and the development of advanced optical atomic clocks. Further applications include LIDAR, coherent communications, frequency synthesis, and precision sensors of strain, motion, and temperature. While all applications benefit from lower frequency noise, many also require a laser that is robust and compact. Here, we introduce a dual-microcavity laser that leverages one chip-integrable silica microresonator to generate tunable 1550 nm laser light via stimulated Brillouin scattering (SBS) and a second microresonator for frequency stabilization of the SBS light. This configuration reduces the fractional frequency noise to $7.8 \times 10^{-14}$ $1/\sqrt{Hz}$ at 10 Hz offset, which is a new regime of noise performance for a microresonator-based laser. Our system also features terahertz tunability and the potential for chip-level integration. We demonstrate the utility of our dual-microcavity laser by performing optical spectroscopy with hertz-level resolution.

**Main Text:**

The stabilization of a continuous-wave laser to an isolated, high-Q Fabry-Perot cavity is the primary technique today for achieving optical frequency references exhibiting the highest levels of spectral purity. In laboratory apparatus, this has resulted in laser linewidths below 1 Hz[1, 2, 3], but these systems are typically comprised of multiple bulk-optic components, such as a fiber or solid-state laser source and a tabletop Fabry-Perot reference cavity that is vibration isolated, thermally insulated, and vacuum housed. In a few cases, these systems have been engineered to allow operation outside the laboratory, where significant vibration and acoustic noise might exist[4, 5]. Narrow linewidths have also been reported by stabilizing lasers to high-resolution interferometers[6] and cryogenic spectral holes[7]. Nonetheless, the size scale of these experiments still prevent their use in many applications, especially those emphasizing portability where robustness, power, weight, and size become primary considerations. However, recent developments of chip-scale high-Q microresonators[8, 9, 10], for both the generation[11, 12, 13, 14] and stabilization[15, 16, 17, 18, 19] of lasers, provide new and largely unexplored opportunities for creating a narrow-linewidth laser source. Taken together with advanced photonic heterogeneous integration[20], it now appears feasible that active laser and electro-optic components could be integrated with high-Q microresonators to realize a frequency-stabilized and narrow-linewidth laser on a silicon chip. It is within this context that we demonstrate a prototype dual-microcavity Brillouin laser that is robust, chip-compatible, and sufficiently low noise for hertz-level applications. We achieve a frequency noise of 200 $Hz^2$/Hz at 10 Hz offset, which corresponds to a single-sideband phase noise of 0 dBc/Hz on the 193 THz carrier.

This reduction in noise brings the dual-microcavity Brillouin laser's operation into a regime below that previously achieved with integrated components[17, 19, 21]. Moreover, a laser source with such spectral purity shows promise for enabling real-world applications. In the case of optical-frequency division (OFD)[22, 23, 24], the conversion of an optical frequency at 1550 nm into an RF frequency at 10 GHz would result in a phase-noise reduction of 86 dB. With ideal frequency division, the achievable phase noise at 10 GHz derived from our stabilized SBS laser becomes -85 dBc/Hz at a 10 Hz offset. This level of phase noise is an order of magnitude lower than that of the highest performing, room temperature 10 GHz commercial oscillators. Through future efforts in the miniaturization of the optical frequency dividers[25, 26, 27, 28, 29], this level of performance can potentially be brought to the system scale of a compact package. Beyond OFD, the dual-microcavity Brillouin laser is also expected to enable greater detection sensitivity in fiber-optic sensors[30], lower bit-error rates in coherent optical communications[31] and finer resolution in spectroscopic measurements[32].

Our narrow-linewidth laser relies on stimulated Brillouin scattering (SBS)[33] lasing in a silica microdisk. Our laser exhibits an excellent white-frequency-noise floor and low optical threshold-power (< 0.1 mW) on a silicon chip[13, 34, 35]. By confining the optical power to a high-Q microcavity, the intracavity power is significantly enhanced while the intracavity losses are minimized. The combination of high circulating optical power and low optical threshold gives rise to large a signal-to-noise ratio, beyond that achievable by conventional lasers. Taken together, these properties allow for low-noise SBS lasing at fractions of a milliwatt in pump power. To enable further applicability as an integrated narrow-linewidth source, two challenges remain for the SBS laser: (1) the ability to efficiently frequency tune the SBS laser, while still retaining a compact system configuration; and (2) the reduction of SBS laser noise at low frequencies, where thermal fluctuations strongly degrade noise performance.

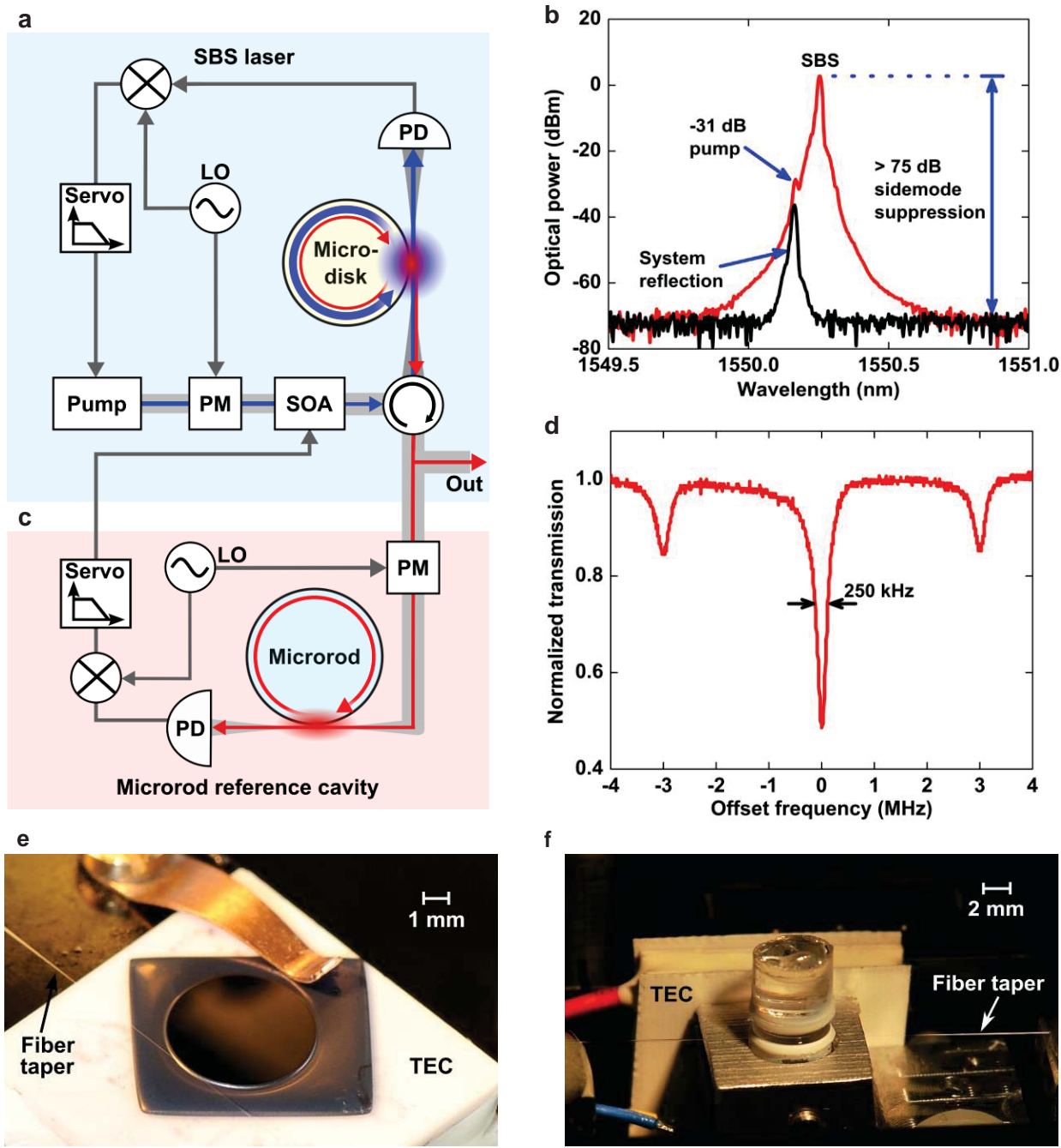

**Figure 1. Diagram and operation of the dual-microcavity narrow-linewidth laser. a**, An integrated planar diode-laser is modulated by a phase modulator (PM) and optically amplified by a SOA before being sent into a microdisk for pumping the generation of a counterpropagating SBS wave. The transmitted signal past the microdisk is detected by a photodetector (PD), which when mixed with a local oscillator (LO), produces an error signal that is used to PDH lock the pump to the peak of a cavity resonance. Through control of the SOA optical output power, the cavity resonances are thermally tuned which results in tuning of the SBS frequency. **b**, Optical spectrum of a 1550.2 nm SBS signal exhibiting a peak power of 1.9 mW and a sidemode suppression greater than 75 dB. The backscattered pump is 31 dB below the SBS signal and at a level 7.8 dB above residual system reflections. **c**, The SBS signal is phase modulated and PDH locked to a stable microrod reference cavity via feedback on the SOA power. **d**, Scan of the SBS laser over a microrod resonance. The linewidth is 250 kHz calibrated from two phase modulation sidebands at ±3 MHz. **e**, Photograph of the microdisk on a silicon chip that is mounted on a thermoelectric cooler (TEC) and coupled to a tapered fiber. **f**, Photograph of the microrod reference cavity.

To address the first challenge, we present a way to frequency tune the SBS laser through control of the interaction between optical power and the local temperature of the optical mode. Our approach only requires control over the pump power, which we provide through the use of an additional chip-scale semiconductor optical amplifier (SOA), thus removing the need for tuning through electro-optic or acousto-optic modulators and additional RF synthesizers. Our approach using a SOA avoids the need to frequency tune the SBS pump laser, which further minimizes noise. The application of our frequency control allows us to directly address the second challenge noted above. Here, we lock the SBS laser to a resonance of a compact high-Q microrod reference[36, 37] exhibiting excellent thermal stability, which results in an improvement of the SBS laser's low-frequency noise by four orders of magnitude.

**SBS generation and tuning**

We generate SBS laser light by pumping a 6-mm-diameter silica microdisk cavity having a loaded Q of $10^8$ (Fig. 1a). The pump laser we use is a commercial integrated planar external-cavity diode laser assembled within a standard 14-pin butterfly package that we couple to the microdisk through a fiber taper. To reduce the transfer of pump noise into the SBS signal, the pump laser is externally phase-modulated and locked to the microdisk resonance. This is accomplished using the Pound-Drever-Hall (PDH) technique[38] with feedback performed on the pump current, which over our range of operation primarily controls the pump frequency. The frequency tuning of the SBS laser is accomplished by sending the pump through a commercial fiber pig-tailed SOA operating in saturation whose output power is controllable via the SOA operating current[39]. Because the local temperature of the optical mode varies with the input power into the microresonator, the position of the cavity resonance is precisely tuned through control of the SOA. Furthermore, since the pump laser is locked to the cavity resonance, the resulting generated SBS signal tunes in unison with the cavity response. In conjunction with the SOA, we have also mounted the microdisk onto a temperature controller to provide temperature tuning of the cavity resonances over a wider frequency range. The optical spectrum corresponding to a 1.9-mW SBS laser is shown in Fig. 1b pumped with a total amplified output of 10 mW.

The sensitivity of the microdisk to thermal fluctuations prevents the SBS laser from reaching its optimum noise performance at low Fourier frequencies (<10 kHz). To improve on the laser's noise, we use our frequency control of the SBS laser to lock its output to a 6-mm-diameter microrod reference cavity (Fig. 1c), whose larger thermal response time significantly increases stability at longer time scales (0.1 ms to 1 s). In order to reduce the impact of environmental perturbations, we temperature control the microrod and enclose it to reduce air currents. The simplicity of our isolation scheme makes it well suited for a compact system implementation. Figure 1d shows a measurement of the SBS laser externally phase-modulated at a rate of 3 MHz and subsequently scanned across a microrod resonance. From this measurement, we estimate the cavity's intrinsic Q to be 0.9 billion. A slight (5 %) linear baseline offset, resulting from the variation in SBS power over the course of the scan, has been removed from Fig. 1d.

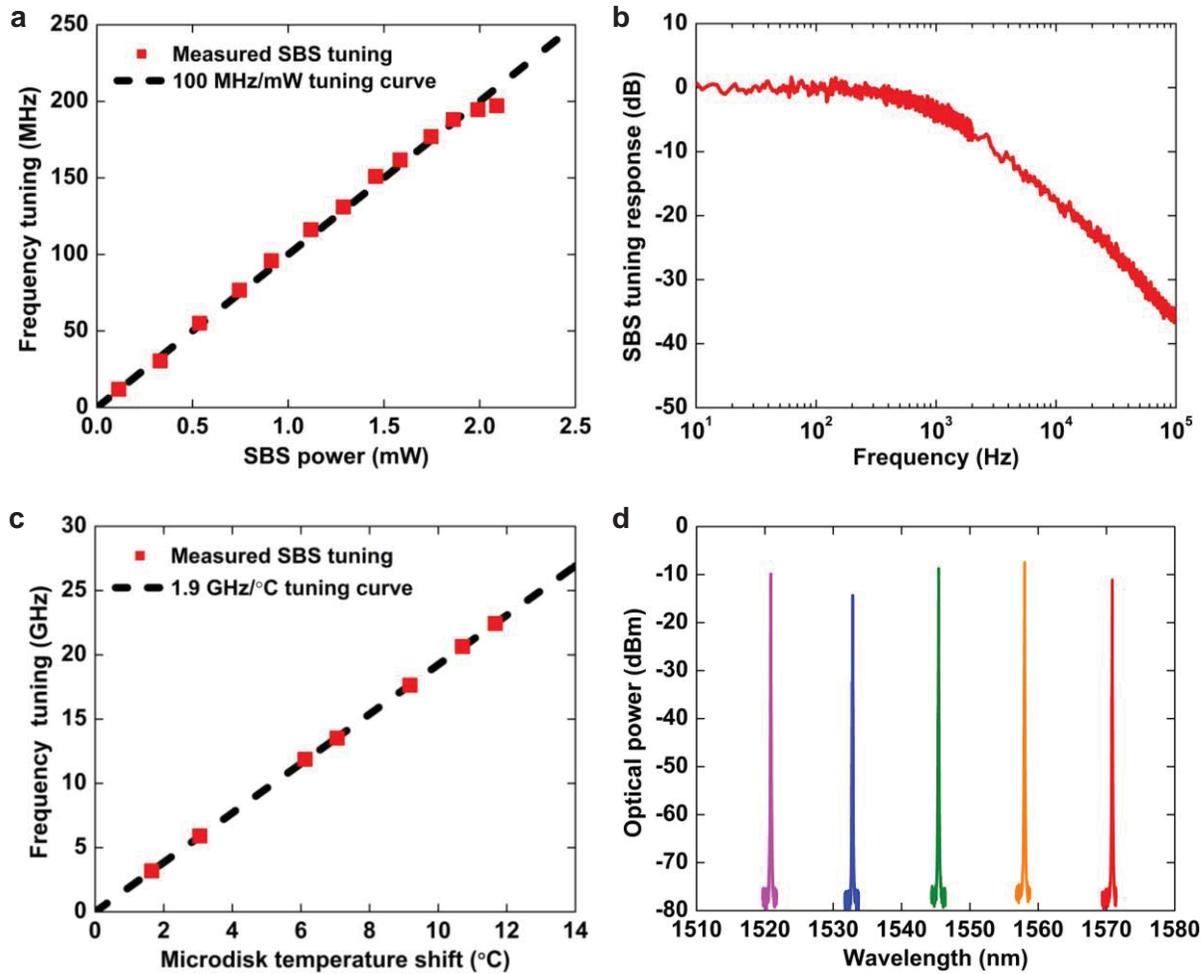

**Figure 2. Demonstration of SBS tuning. a**, Graph of the measured change in frequency of the SBS laser versus its power for tuning via SOA control. The fitted response follows a tuning rate of 100 MHz per milliwatt of SBS power **b**, Frequency response achieved by the SOA tuning method. The 3-dB bandwidth is ~850 Hz approximately corresponding to the thermal response rate of the microdisk. **c**, Graph of the measured change in frequency of the SBS laser for controlled shifts in the microdisk temperature over a total tuning range of 22.4 GHz. The fitted response follows a tuning rate of 1.9 GHz per °C of temperature change. **d**, SBS optical spectra for five operating points when the pump laser is tuned across a range of 50 nm.

Our approach to the frequency control of the SBS laser accommodates multiple tuning ranges. For tuning on a finer scale, the resonator response to power variations of the optical mode allows for frequency shifts in the range of megahertz. Figure 2a shows the tuning of the SBS laser when the SOA output is scanned, achieving SBS powers in the range of 0 and 2.1 mW. The rate of tuning is 100 MHz frequency shift per milliwatt of SBS power with a total tuning range of 200 MHz. Beyond 200 MHz, the SBS laser tuning reaches the limit where the pump laser can stay locked to the cavity resonance, set by the bounds of the servo control. By using a pump laser that can be tuned further, the SBS tuning range can be extended. Figure 2b shows a measurement of the frequency response corresponding to this tuning method, which depicts the conversion of an amplitude-modulated input into a frequency-modulated output. The frequency response exhibits a bandwidth of ~850 Hz and provides a measure of the system's rate of thermal response. We note that our approach to tuning the SBS laser here self-adaptively adjusts to the noise of the SBS laser, and is therefore ideally suited for locking the SBS to the microrod reference cavity. At lower frequencies where the SBS noise is high due to thermal fluctuations, the tuning response is large providing ample levels of servo gain to correct for frequency errors. At higher frequencies where the SBS noise is low, the tuning response cuts off minimizing the effect of noise introduced by the servo.

In many cases, a larger tuning range is desirable so that the laser can first be positioned to a location of interest before being subsequently scanned across a narrower window. We achieve this range through direct control of the microdisk temperature, which enables the cavity resonances to be tuned across multiple gigahertz. Figure 2c shows the change in frequency of the SBS laser when the microdisk temperature is varied. The SBS frequency changes linearly with temperature at a tuning rate of 1.9 GHz/°C. We note that due to the previously mentioned limitations of the pump's locking range, our feedback onto the pump current breaks down once the tuning exceeds ~200 MHz. Hence, the measurements of Fig. 2c are performed by manually tuning the temperature of the pump laser so as to follow the shift in the microdisk resonance before locking the pump at the new desired location. As before, the use of a pump laser capable of being tuned farther would allow the SBS frequency to be continuously tunable over a broader range. After shifting the microdisk temperature by 11.7 °C, we achieve a SBS tuning range of 22.4 GHz.

Since the modes of the microdisk are periodically spaced in frequency, we can alternatively tune the pump laser but keep the microdisk modes fixed consequently using different cavity resonances for SBS generation. With this method, the tuning range of the SBS laser becomes primarily limited by the tuning capabilities of the pump laser. Figure 2d shows the optical spectrum of the SBS laser for five different points when the pump laser is scanned across its operation range from 1520 nm to 1570 nm. In this case, the total SBS tuning is larger (50 nm) but is no longer continuously tunable across the frequency scan. This method of SBS tuning is therefore most suitable for moving the pump to a convenient frequency range before applying the techniques of Figs. 2a and c for finer control of SBS frequency. Note, however, that the SBS gain peak shifts proportional to the pump frequency, such that with still greater tuning the phase-matching requirements of the SBS laser are no longer satisfied without changing the microdisk dimensions[34]. Nonetheless, because the full tuning range potentially encompasses the entire transparency window of silica, the SBS nonlinearity can be further used to enable narrow-linewidth lasing in the near-infrared or visible range. To this end, low-noise SBS lasing at a wavelength of 1064 nm has been recently demonstrated with microdisks that are ≈4 mm in diameter[35].

**Dual-microcavity SBS laser characterization**

The ability to control the SBS laser's frequency enables the laser to be used for a wide array of applications. Here, we use this frequency control to lock the laser to a microrod cavity, whose stability at lower frequencies significantly reduces the noise exhibited by the SBS laser. We quantify the locked laser's noise through two techniques. In one method, we use a Mach-Zehnder interferometer setup with delay lengths varying between 15 m and 300 m to quantify the SBS laser's frequency fluctuations. In the second method, we directly heterodyne the laser with a sub-hertz linewidth cavity-stabilized Er fiber laser and subsequently detect the fluctuations using a frequency-to-voltage converter. Figure 3a shows the measured frequency noise illustrating the progression in noise reduction from our pump laser to the SBS laser and finally to the microrod stabilized SBS laser. At high frequencies, the SBS laser improves on the pump noise by over two orders of magnitude reaching a noise floor below 1 Hz$^2$/Hz. However, at lower frequencies (< 1 kHz), the fluctuations of the SBS laser drives its noise nearly to the level of the pump laser's noise. By locking the SBS laser to the microrod cavity, the noise at lower frequencies is improved, reaching a noise reduction by four orders of magnitude at 10 Hz. Our servo cuts off near 10 kHz (see Fig. 2b) and thus adds negligible noise to the SBS laser at higher frequencies. Therefore the locked SBS laser noise is low at all frequencies in Fig. 3a, unlike typical cavity-stabilized lasers where the noise strongly degrades beyond the locking bandwidth. Although not shown, we note for reference that the corresponding single-sideband phase noise values at 10 Hz, 1 kHz, and 100 kHz are 0 dBc/Hz, -55 dBc/Hz, and -104 dBc/Hz, respectively.

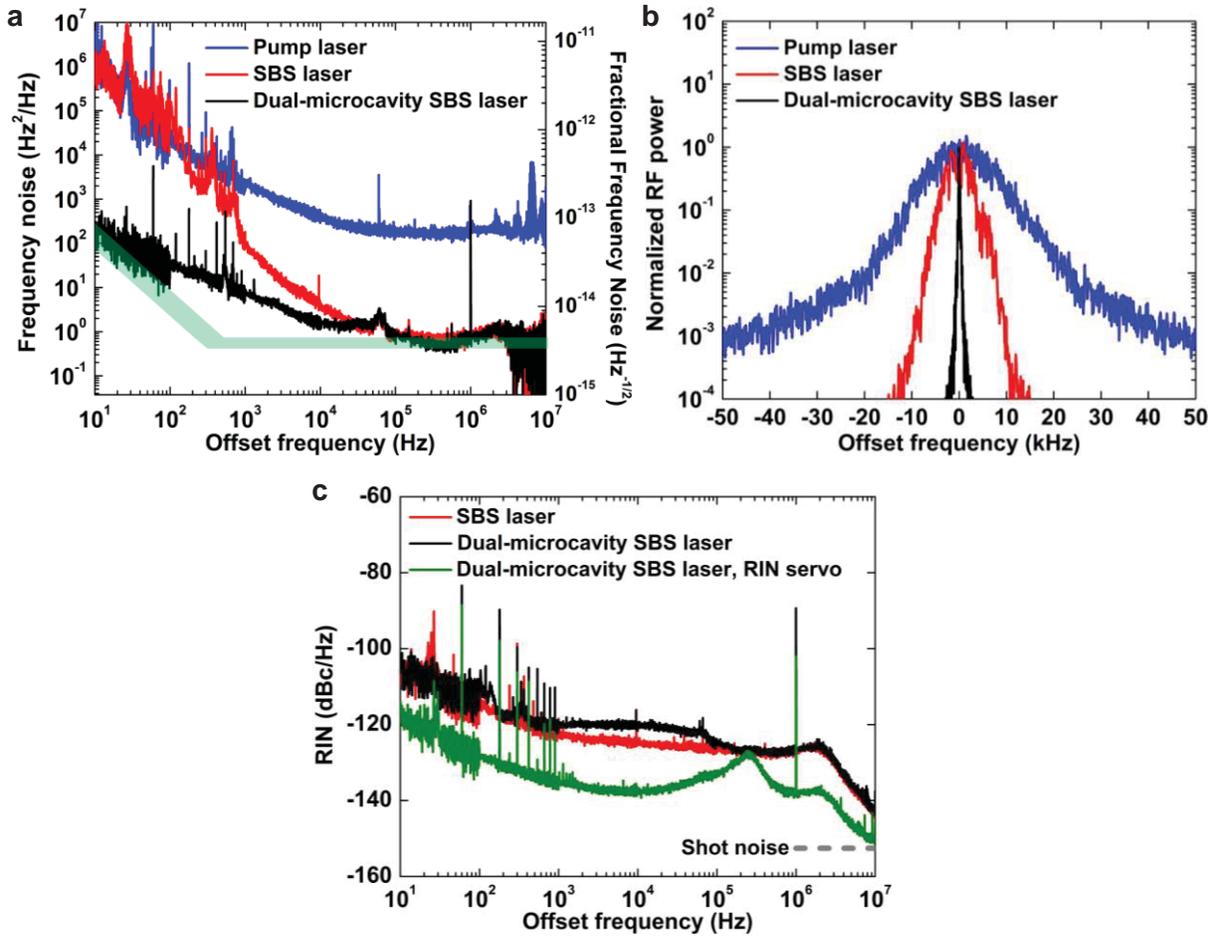

**Figure 3. Experimental measurements of the SBS laser. a**, Frequency noise spectra of the pump (blue), SBS (red), and dual-microcavity SBS (black) lasers. The SBS laser improves on the pump noise at high offset frequencies but suffers from noise fluctuations at low frequencies. When locked to the microrod, the noise at low offset frequencies becomes significantly reduced. The shaded region indicates the uncertainty window corresponding to theoretical calculations of the dual-microcavity laser's intrinsic noise. This noise results from a combination of the reference cavity's thermorefractive noise at lower offsets and the SBS laser's fundamental white-frequency noise at higher offsets. **b**, RF spectrum illustrating the linewidth reduction from the pump laser to the SBS laser and finally to the locked SBS laser. **c**, RIN spectra of the SBS and locked SBS lasers showing a slight increase in RIN after locking to the microrod. By servoing the intensity noise, the locked SBS RIN (green) can be reduced.

Our calculations reveal that the measured frequency noise may be approaching the level of the microrod reference cavity's thermorefractive noise[15, 40] at lower offset frequencies (shaded region of Fig. 3a). For these frequencies, the bounds of the shaded region correspond to uncertainties in the microrod mode-field diameter, which we estimated to be between 100 and 200 μm. At higher offsets, the noise reaches a floor set by the fundamental limits of the SBS process. For the calculations of the white-noise floor, we bounded the microdisk effective mode area to be between 25 μm$^2$ and 150 μm$^2$ and varied the external coupling ratio to maintain a constant output power of 1.9 mW. We find from Fig. 3a that the measured dual-microcavity SBS laser noise is accurately captured by the calculated white noise floor for frequencies above 100 kHz.

Figure 3b shows the RF spectra corresponding to the pump, SBS, and locked SBS lasers measured against the sub-hertz linewidth cavity-stabilized reference laser. For a spectrum analyzer sweep time between 0.1 and 0.2 s, we measure the half-power width of the pump laser to be 7.9 kHz, while we measure the width of the SBS laser to be narrower at 3.3 kHz limited by low-frequency noise. In comparison, the locked SBS laser exhibits a half-power width of 95 Hz. The fraction of the pump and unstabilized SBS laser powers residing within a 95-Hz window is 1.1 % and 2.8 %, respectively and thus illustrates the improvement gained by locking to the microrod. In order to estimate the linewidth of the locked SBS laser, we convert Fig. 3a to phase noise and integrate from higher frequencies towards lower frequencies[41]. This estimate yields a laser linewidth of 87 Hz, which is in good agreement with our measured half-power width of 95 Hz.

The relative intensity noises (RIN) of the SBS and locked SBS lasers (Fig. 3c) are measured by direct photodetection of the optical signal. Between 1 kHz to 100 kHz, the frequency stabilization of the SBS laser slightly adds to the total level of intensity noise. At 2 MHz, the SBS laser exhibits a damped relaxation oscillation, beyond which the noise decays to the level of shot noise. We improve the SBS laser RIN by amplifying the SBS output using a SOA and subsequently locking the intensity through control of the SOA current. By operating the SOA in saturation, the SBS RIN is further damped for frequencies up to 10 GHz. Because the SBS power is generally maintained to be relatively low in order to prevent the cascaded generation of higher order SBS waves, the SOA serves the dual purpose of also increasing the laser's usable output power. Finally, since our method of frequency tuning via pump power causes the SBS power to also vary, this intensity servo helps in stabilizing the optical power against changes in the operating point. Figure 3c shows the RIN of the locked SBS laser sent through the intensity servo with 100 kHz of loop bandwidth. Over most of the frequency range, the RIN decreases by over an order of magnitude matching the level of RIN exhibited by the amplified pump laser (not shown). For this measurement, the total SBS laser output power is 15 mW but can be further increased to 30 mW with even larger SOA biases. We note that the addition of the SOA slightly increases the SBS laser's frequency noise beyond 1 MHz but negligibly contributes to noise at lower frequencies.

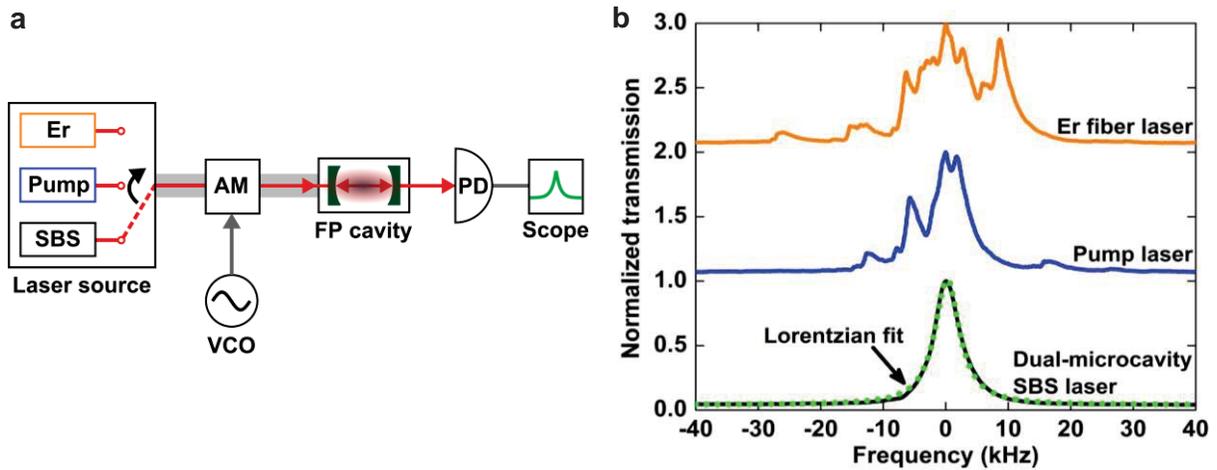

**Figure 4. Laser spectroscopy of a narrow-linewidth bulk Fabry-Perot cavity. a**, Three lasers are individually sent through an amplitude modulator (AM) to generate sidebands that can be scanned by a VCO (voltage controlled oscillator). The laser sidebands are swept across a Fabry-Perot (FP) cavity resonance exhibiting a linewidth of ~4 kHz. The resulting cavity transmission is photodetected and subsequently recorded by an oscilloscope. **b**, Scan of the dual-microcavity SBS laser over the cavity resonance along with its corresponding Lorentzian fit. The scans corresponding to the pump laser and a commercial narrow-linewidth Er fiber laser are also shown for comparison. The cavity resonance is only accurately resolved in the SBS laser trace.

### Laser spectroscopy of a high-Q cavity

The dual-microcavity SBS laser's spectral purity enables high-resolution measurements of spectrally-narrow features. We demonstrate this capability by scanning the locked SBS laser over a narrow (~4 kHz) resonance of a high-Q Fabry-Perot cavity (Fig. 4a). Here we use a separate electro-optic modulator and synthesizer to scan the SBS frequency. Figure 4b shows the locked SBS scan over the cavity resonance along with comparison scans performed using the pump laser and a commercial low-noise Er fiber laser. We ensure that uniformity is maintained in the scanning conditions for all three lasers. The trace of the Er laser's transmission gives indication of multiple peaks due to the laser's frequency noise sampling the cavity resonance over the duration of the scan. The resolution of this lineshape is improved with the pump laser scan; however, noticeable distortion of the resonance is still present. In contrast, the cavity lineshape is accurately resolved during the scan with the locked SBS laser. The ratio of noise to signal in Fig. 4b is below $1\times10^{-3}$ at the point of half transmission. Assuming systematic uncertainties can be accounted for, the demonstrated signal to noise would enable resolution of the center frequency to a level below 4 Hz.

### Conclusions

In summary, we have demonstrated a dual-microcavity Brillouin laser that features the powerful combination of tunability and fractional frequency noise below the level of $10^{-13}$ $1/\sqrt{Hz}$. Our method of tuning the SBS laser provides a path towards enabling the laser's use for many applications where frequency control is necessary, and this approach may be suitable for use in other microcavity based systems (e.g., the microresonator frequency combs). The present system, which already employs full fiber integration between the components, is a prototype for future developments that could include chip-level integration. In addition to the SBS disk resonator, the functionality of all of the components in our system has been demonstrated in photonic integrated circuits[20]. This includes reference cavities[19] and chip-based tunable filters[42] that could provide the capacity to extract the SBS signal from the pump. Finally, we note that while the present frequency noise level is approaching thermorefractive limits, the use of a larger mode-volume reference cavity and an SBS laser with higher circulating power provide a route to still improved frequency noise in our dual-microcavity system.

## Methods:

### SBS laser tuning characterization

We characterize the tuning range of the SBS laser by beating the laser against a Er fiber frequency comb stabilized to a hydrogen maser. By monitoring the shift of the beat note, we directly measure the shift of the SBS frequency. This method accurately determines the SBS tuning for frequency shifts of a few hundred megahertz. To detect larger tuning ranges, we heterodyne the SBS laser against a cavity-stabilized Er fiber laser. This beat note is detected by a fast photodiode and monitored on a RF spectrum analyzer. The tuning of the SBS laser on the nanometer scale is found by direct detection of the SBS signal on an optical spectrum analyzer.

### SBS laser frequency noise measurement

To measure the frequency noise of the dual-microcavity SBS laser, we first send its signal output through a Mach-Zehnder interferometer setup with varying delay lengths between 15 m and 300 m. The use of varying delays allows us to trade off measurement sensitivity and measurement range. By piecing together the ensemble of collected data, the SBS laser's frequency noise can be determined over a 10 MHz range. At low frequencies, this technique becomes sensitive to vibration, so we instead measure the frequency noise by heterodyning the laser against a cavity-stabilized Er fiber laser. The resulting beat signal occurs near 13.5 GHz, which we subsequently mix to 30 MHz via a RF synthesizer. The 30 MHz signal is then sent through a frequency-to-voltage converter and finally into a FFT analyzer for processing of its frequency noise. We note that this second measurement technique provides excellent sensitivity at low frequencies but suffers from reduced sensitivity at higher frequencies where the Er fiber laser loses servo control. Thus the two techniques can be used together to provide a full characterization of the laser frequency noise.

**Acknowledgements:** We thank Dr. Nathan Newbury and Dr. Jeffrey Sherman for their comments on this manuscript. We also thank Dr. Pascal Del'Haye and Dr. Aurélien Coillet for useful discussions. This work is funded by NIST and the DARPA PULSE program. W.L. acknowledges support from the NRC/NAS. This work is a contribution of the US Government and is not subject to copyright in the US.

**Author Contributions:** W.L., S.B.P., and S.A.D. conceived the experiments. H.L. and K.J.V. fabricated and provided the microdisk resonators. W.L. developed and characterized the SBS laser source. A.A.S.G. built the microrod reference cavity. Both W.L. and A.A.S.G. constructed and characterized the dual-microcavity SBS laser. F.N.B, D.C.C., and F.J.Q. built and developed stabilized CW laser sources and optical frequency combs that assisted with measurements. All authors contributed to the manuscript preparation.

**Competing financial interests:** The authors declare no competing financial interests.

**Corresponding Authors:** Correspondence to W. Loh